\documentclass[pra,onecolumn,showpacs,preprintnumbers,amsmath,amssymb,superscriptaddress]{revtex4}

\usepackage{graphicx}
\usepackage{dcolumn}
\usepackage{bm}
\usepackage{amsmath}
\usepackage{latexsym}

\input{tcilatex}

\begin{document}

\title{The stationary solutions of G-P equations in double square well}
\author{WeiDong LI}
\affiliation{Department of Physics and Institute of Theoretical Physics, Shanxi\\
University, Taiyuan 030006, China}
\affiliation{INFM-BEC and Dipartimento di Fisica, Universita di Trento, 1-38050 Povo}

\begin{abstract}
We present analytical stationary solutions for the
Gross-Pitaevskii equation (GPE) of a Bose-Einstein condensate
(BECs) trapped in a double-well potential. These solutions are
compared with those described by [Mahmud et al., PRA \textbf{66},
063607 (2002)]. In particular, we provide further evidence that
symmetry preserving stationary solutions can be reduced to the
eigenstates of the corresponding linear Schr\"{o}dinger equation.
Moreover, we have found that the symmetry breaking solutions can
emerge not only from bifurcations, but also from isolated points
in the chemical potential-nonlinear interaction diagram. We also
have found that there are some moving nodes in the symmetry
breaking solutions.
\end{abstract}

 \maketitle


\section{Introduction}

The Gross-Pitaevskii equation (GPE) describes many features of
Bose-Einstein condensates (BECs) of dilute atomic gases in an
external potential at zero temperature \cite{sandro01}. The
properties of the ground state of the GPE with external potential
have been extensively studied; and many interesting phenomena have
been reported. Of special interest is that the mean filed
interaction profoundly modifies the density profiles and the
stability of the ground state \cite{sandro01}. Recently, the
properties of the nonground-state stationary solutions of the GPE
have attracted more attention both theoretically and
experimentally \cite{kutz01,presilla01,philip01}. For example, the
dark solitons have been created in the atomic gases with positive
scattering length by phase engineering optical techniques \cite
{philip01,burger01,niu}.

Based on \cite{presilla01}, one can classify the stationary
solutions of GPE as symmetry preserving and symmetry breaking
solutions. Whereas the vortices and solitons observed in
experiments \cite{philip01,burger01,salomon01,hulet01} are
symmetry preserving solutions, which can be reduced to the
eigenstates of the corresponding linear Schr\"{o}dinger equations,
the macroscopic quantum self-trapping state in the two states
\cite{smerzi01,smerzi02,gati01,gati02} and non-Bloch state
\cite{pethick01,smerzi03} in periodical potential are symmetry
breaking solutions, and they can not be reduced to the eigenstates
of the corresponding linear Schr\"{o}dinger equations.

The system of GPE with double well is a good system to investigate
the properties of the stationary solutions of GPE and also good
for studying the special nonlinear dynamics, for example, the
nonlinear self-trapping effect which was predicted theoretically
in 1997 \cite{smerzi01} and realized experimentally
\cite{gati01,gati02,gati03} last year. It has been shown that
there are stationary solutions, which are either symmetry
preserving or symmetry breaking, both numerically and analytically
in this system \cite{kutz01,presilla01} . With the help of the
double square well \cite{kutz01}, which allows one kind of
analytical solution for GPE, Reinhardt and his collaborators have
confirmed the numerical calculation for GPE with double-well trap
\cite{presilla01}. This provides one possible way to investigate
the properties of the stationary solution for GPE with double
well. It is interesting to know how the stationary solutions
change as the nonlinear interaction increases or decreases, how
the symmetry breaking stationary solutions emerge and how many
kinds of the stationary solutions there are for the fixed
nonlinear interaction.

In this paper, we present new stationary analytical solutions for
GP equation with double square well. Compared with the solutions
of \cite {kutz01}, ours can be reduced to the eigenstates of the
corresponding linear Schr\"{o}dinger equations. It is this feature
which enables us to understand the above-mentioned interesting
problems concerning the stationary solutions for GPE. It allows us
to obtain the critical nonlinear interaction value, over which the
symmetry breaking solution emerges. And from the relation of the
profiles of the stationary solutions and chemical potentials with
the nonlinear interaction, we can directly see the means of the
symmetry and symmetry-breaking and different stationary solutions
for special nonlinear interaction.

\section{Model and the solutions}

Considering that the BECs of dilute atomic gases are confined by a
very anisotropic harmonic potential ($\omega_{\bot} \gg
\omega_{x}$, where $\omega_{\bot,x}$ are the confined frequency in
the y-z space and x direction) and that the BECs are loaded in a
double square well in the weak confined direction (x direction),
then the dynamics of this system is governed by the 1-D G-P
equation
\begin{equation}
i\hbar \frac{\partial \Psi \left( x,t\right) }{\partial t}=\left( -\frac{%
\hbar ^2}{2m}\frac{\partial ^2}{\partial x^2}+V\left( x\right) +g_0\left|
\Psi \left( x,t\right) \right| ^2\right) \Psi \left( x,t\right)  \label{g1}
\end{equation}
where $g_0=\frac{4\pi \hbar ^2a}m$ is the 1-D reduced nonlinear
interaction constant. The potential is of the form
\[
V\left( x\right) =\left\{
\begin{array}{lcl}
\infty & \quad & |x|\ge a \\
0 &  & b<|x|<a\qquad \quad (a=1/2,V_0>0) \\
V_0 &  & |x|<b,%
\end{array}
\right.
\]
The stationary solution can be written as $\psi \left( x\right)
=r\left( x\right) \exp \left( -i\frac \mu \hbar t\right) $ and
after re-scaling the equations, we arrive at the equation for
$r\left( x\right) $
\begin{equation}
\mu r\left( x\right) =-\frac{\partial ^2}{\partial x^2}r\left( x\right)
+V\left( x\right) r\left( x\right) +\eta r\left( x\right) ^3  \label{str}
\end{equation}
where $\eta =N_0g_0\frac{2mL^2}{\hbar ^2}$ and $N_0$ is the total
number of the atom. The energy and the potential are measured in
unit of $\frac{\hbar ^2}{2mL^2}$, and $L$ is the length of the
total space (here $L=2a$). Due to the double well case, the
stationary solution is just the real function, so we have assumed
that our solution $r\left( x\right) $ is real function. As in Ref.
\cite{kutz01}, the solution of (\ref{str}) can be written in terms
of the Jacobi elliptical function.

Generally, we have two different solutions depending on the
relation of the chemical potential and the barrier height. First
for $\mu >V_0$

\begin{equation}
r_1\left( x\right) =A \cdot \text{SN}\left( Kx+\delta ,n_1\right)
, \label{f1}
\end{equation}
where
\begin{equation}
n_1=\frac{A^2}{2K^2}\eta ,\qquad \mu =K^2+V_0+\frac{A^2}2\eta .  \label{f2}
\end{equation}
This solution is also valid for $V_0=0$, which corresponds to the region of $%
b<\left| x\right| <a$. It is easy to check that when $\eta =0$, our solution
is reduced to $r_1\left( x\right) =A\sin \left( Kx+\delta \right) ,$where $%
n_1=0$, $\mu =K^2+V_0.$ This is nothing but the eigenstates of
linear Shr\"{o}dinger equations for $\mu >V_0$. We have two
different Jacobi functions for $\mu <V_0$, corresponding to the
region of $\left| x\right| <b$. To the case with one node in the
barrier region, the solution is

\begin{equation}
r_2\left( x\right) =B \cdot \text{SC}\left( Qx+\gamma ,n_2\right)
, \label{b1}
\end{equation}
\begin{equation}
n_2=1-\frac{B^2}{2Q^2}\eta ,\qquad \mu =V_0-Q^2-\frac{B^2}2\eta .  \label{b4}
\end{equation}

\begin{table}[hb]
\caption{ The modular transformation of Jacobi elliptic function}
\label{ta}\centering
\begin{tabular}{|l|l|l|l|l|l|}
\hline\hline & $m<0$ & $0<m<1$ & $m>1$ & $m=0$ & $m=1$ \\
\hline
SN$\left( u|m\right) $ & $\sqrt{m_1}$SD$\left( u\sqrt{1-m}|m_1\right) $ & SN$%
\left( u|m\right) $ & $\sqrt{m_2}$SN$\left( \frac u{m_2}|m_2\right) $ & $%
\sin \left( u\right) $ & $\tanh \left( u\right) $ \\ \hline
SC$\left( u|m\right) $ & $\sqrt{m_1}$SC$\left( u\sqrt{1-m}|m_1\right) $ & SC$%
\left( u|m\right) $ & $\sqrt{m_2}$SD$\left( \frac u{m_2}|m_2\right) $ & $%
\tan \left( u\right) $ & $\sinh \left( u\right) $ \\ \hline
CN$\left( u|m\right) $ & CD$\left( u\sqrt{1-m}|m_1\right) $ &
CN$\left( u|m\right) $ & DN$\left( \frac u{m_2}|m_2\right) $ &
$\cos \left( u\right) $ & $\sec h\left( u\right) $ \\ \hline
DN$\left( u|m\right) $ & ND$\left( u\sqrt{1-m}|m_1\right) $ &
DN$\left( u|m\right) $ & CN$\left( \frac u{m_2}|m_2\right) $ & $1$
& $\sec h\left( u\right) $ \\ \hline
\end{tabular}
\end{table}

but to the case without node,
\begin{equation}
r_2\left( x\right) =B \cdot \text{NC}\left( Qx+\gamma ,n_2\right)
, \label{b2}
\end{equation}
\begin{equation}
n_2=1-\frac{B^2}{2Q^2}\eta ,\qquad \mu =V_0-Q^2+B^2\eta .  \label{b3}
\end{equation}

Same as the solution (\ref{f1}) in $\eta =0$, our solutions are reduced to $%
r_2\left( x\right) =B\sinh \left( Qx+\gamma \right) $ for (\ref{b1}) and $%
r_2\left( x\right) =B\cosh \left( Qx+\gamma \right) $ for (\ref{b2}), where $%
n_2=1$, $\mu =V_0-Q^2$. It is interesting to note that those two
solutions are precisely identical with the eigenstates of linear
Shr\"{o}dinger equations for one node or no node within the
barrier.

Please note here that we do not restrict the value of $n_1$ from
$0$ to $1$ as is usually used in Jacobi elliptic function. But
this problem could be solved by the modular transformation table
(see table \ref{ta}, where $m_1=\frac m{1+m}$, $m_2=\frac 1m$.)
\cite{byrd01}. The Jacobi elliptical function sc and nc are
constructed from the Jacobi elliptical SN, CN and DN (see (table
\ref{tb}) or \cite{byrd01}).

\section{Symmetry preserving and symmetry breaking solutions}

As mentioned in Introduction, we have two different kinds of
stationary solutions depending on whether the stationary solution
has its linear counterpart \cite{presilla01}. The symmetry
preserving solution has the linear counterpart, as it could be
reduced to the eigenstates of the corresponding linear
Schr\"{o}dinger equations. But the symmetry breaking solution
could not be reduced, therefore it does not have the linear
counterpart. Such being the case, we are required to find the
stationary solutions of Eq. (\ref{g1}). It is worthwhile to note
that our method is just the usual one in linear case.

\begin{table}[h]
\caption{Other Jacobi elliptic functions}
\label{tb}\centering
\begin{tabular}{|l|l|l|}
\hline\hline
NS$\left( u|m\right) \equiv \frac 1{\text{SN}\left( u|m\right) }$ & SC$%
\left( u|m\right) \equiv \frac{\text{SN}\left( u|m\right)
}{\text{CN}\left( u|m\right) }$ & SD$\left( u|m\right) \equiv
\frac{\text{SN}\left( u|m\right) }{\text{DN}\left( u|m\right) }$
\\ \hline
NC$\left( u|m\right) \equiv \frac 1{\text{CN}\left( u|m\right) }$ & CS$%
\left( u|m\right) \equiv \frac{\text{CN}\left( u|m\right)
}{\text{SN}\left( u|m\right) }$ & CD$\left( u|m\right) \equiv
\frac{\text{CN}\left( u|m\right) }{\text{DN}\left( u|m\right) }$
\\ \hline
ND$\left( u|m\right) \equiv \frac 1{\text{DN}\left( u|m\right) }$ & DS$%
\left( u|m\right) \equiv \frac{\text{DN}\left( u|m\right)
}{\text{SN}\left( u|m\right) }$ & DC$\left( u|m\right) \equiv
\frac{\text{DN}\left( u|m\right) }{\text{CN}\left( u|m\right) }$
\\
\hline
\end{tabular}
\end{table}

With the help of (\ref{f1}), (\ref{b1}) and (\ref{b2}), solutions
in the three regions can be written in the form
\begin{eqnarray}
f_1\left( x\right) &=&A_1 \cdot \text{SN}\left(
K_1(x+a),n_1\right) ,\qquad -a<x<-b
\nonumber \\
f_2\left( x\right) &=&B \cdot \text{NC}\left( Q(x+\gamma
),m\right) ,\qquad \qquad
\left| x\right| <b \\
f_3\left( x\right) &=&A_2 \cdot \text{SN}\left(
K_2(x-a),n_2\right) ,\qquad b<x<a \nonumber
\end{eqnarray}
for the case without node inside the barrier. For the case with
one node inside the barrier, we have
\begin{eqnarray}
f_1\left( x\right) &=&A_1 \cdot \text{SN}\left(
K_1(x+a),n_1\right) ,\qquad -a<x<-b
\nonumber \\
f_2\left( x\right) &=&B \cdot \text{SC}\left( Q(x+\gamma
),m\right) ,\qquad \qquad
\left| x\right| <b \\
f_3\left( x\right) &=&-A_2 \cdot \text{SN}\left(
K_2(x-a),n_2\right) .\qquad b<x<a \nonumber
\end{eqnarray}

We have considered the fact that the solutions vanish on and
outside the potential $\left| x\right| \geq a$. To fix the
parameters $A_1$, $A_2$, $K_1$, $K_2$, $Q$, $B$ and $\gamma $, we
need the continuity
\begin{eqnarray}\label{condi1}
f_1\left( -b\right) &=&f_2\left( -b\right) ,  \nonumber \\
f_1^{\prime }\left( -b\right) &=&f_2^{\prime }\left( -b\right) ,  \nonumber
\\
f_2\left( b\right) &=&f_3\left( b\right) , \\
f_2^{\prime }\left( b\right) &=&f_3^{\prime }\left( b\right) ,  \nonumber
\end{eqnarray}
and normalization conditions
\begin{equation}\label{condi2}
\int_{-a}^{-b}\left| f_1^2\left( x\right) \right|
dx+\int_{-b}^b\left| f_2^2\left( x\right) \right|
dx+\int_b^a\left| f_3^2\left( x\right) \right| dx=1,
\end{equation}
and we need that the chemical potential is the same in different
regions ($\mu _1=\mu _2=\mu
_3$). The definition of the chemical potential $\mu $ can be found in Eqs.(%
\ref{f2}), Eqs.(\ref{b4}) and Eqs(\ref{b3}).

\begin{figure}[tbp]
\includegraphics[width=3.2 in]{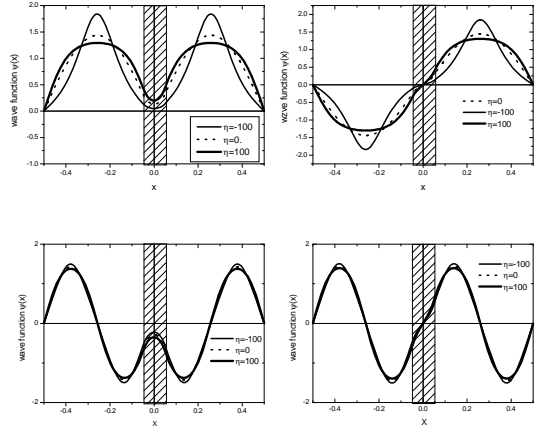}
\vspace{-0.8 in} \centering \caption{The symmetry preserving
solutions of G-P equation} \label{fig1}
\end{figure}

In Fig.1 we present the first four symmetric and antisymmetric
solutions for $ -100<\eta <100$. It is clear that all of these
solutions can be reduced to eigenstates of the corresponding
linear Schr\"{o}dinger equations ($ \eta =0$). All of these
solutions are symmetry preserving solutions. A barrier height of
$V_0=1000$, barrier width $2b=0.1$, well width $2a=1$ are used all
through this paper. The shaded part represents the barrier region
in all the figures. It is hence easy to see that the effect of the
nonlinear interaction on the profile of the wave function of the
high energy state is smaller than that in the case of the low
energy state. This situation is the same as in the lattice case
\cite{smerzi03}. In Fig.4, the chemical potential have been
plotted as the function of the nonlinear interaction. From Fig.4,
one can read that the chemical potential difference between the
ground state and the first excited state increases with the
increase of the strength of the nonlinear interaction. This seems
to suggest that the positive nonlinear interaction enhances
tunneling effect but the negative nonlinear interaction eliminates
it. When the nonlinear interaction is very large and negative (
$\eta _c\sim -54.5$), the symmetrical properties of the ``ground''
state change from the symmetry to the anti-symmetry
\cite{kutz01,presilla01}. In fact, when the nonlinear interaction
is negative and large, the symmetry preserving solution is not the
ground state. But this will be discussed in the later part of the
paper.

\begin{figure}[tbp]
\includegraphics[width=3.2 in]{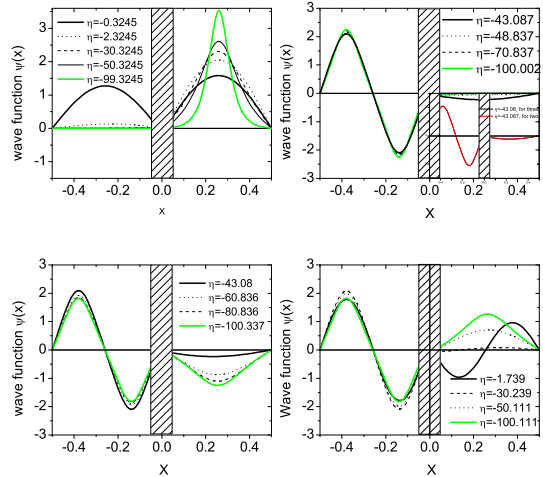}
\centering \vspace{-0.0 in} \caption{(Color online) The symmetry
breaking solutions of G-P equation for positive nonlinear
interaction} \label{fig2}
\end{figure}

In Fig.2 we present the first four symmetry breaking solutions for
positive nonlinear interaction up to $100$. When $\eta =100$,
these four solutions are exactly the one presented in
\cite{kutz01}. In \cite{kutz01}, they also show the profile of the
first kind of solution in Fig.2 for $\eta =15$, $ 30 $, $50$,
$100$. But here our results show that this solution emerges from
the bifurcations of the chemical potential of the first excited
state at $\eta _c\sim 0.32$ (See Fig.4). Some more detailed
behavior can be read from Fig.2. This solution has been predicted
in \cite{smerzi02}. Its chemical potential is larger than that of
the first excited state. It is easy to see that the node in this
solution is moving as the nonlinear interaction increases. From
$0.32$ to $5.563$, the node is displacing from within the barrier
to the right well. The fourth solution emerges from another
bifurcation ($\eta _c\sim 1.48$), at which the chemical potential
is the same as the fourth symmetry preserving solution (the third
excited state). For this solution, there is one fixed node and two
moving nodes. Due to the increase of the nonlinear interaction,
one of the moving nodes moves from the barrier region into the
well. Finally, there are two moving nodes in the right well and
one fixed nodes in the left well. The second and third solutions
in Fig.2 belong to a new kind of symmetry breaking solution. They
occur from the same isolated point ($\eta _c\sim 51.284$) in the
diagram of the chemical potential and the nonlinear interaction.
Here an isolated point is one point whose chemical potential (or
eigenvalue) is not the same as the neighboring points in the
diagram of the chemical potential and nonlinear interaction,
whether it is the symmetry preserving or symmetry breaking
solution. Therefore their profiles are modified completely
differently with the increase of the nonlinear interaction. The
second one has one fixed node and one moving node while the third
one has two moving nodes.

The self-trapping effect as predicted in 1997 \cite{smerzi02} was
experimentally realized in \cite{gati01,gati02,gati03}. This
dynamical effect can be understood very well based on the two mode
approximation \cite{smerzi02}. In our case, due to the zero phase
difference between the two well wave functions, the condition for
the self-trapping can be written as $\Lambda =
2(1+\sqrt{1-\Lambda^2})/\Lambda^2$. It is easy to check that all
of the symmetry breaking solutions satisfy this condition. We show
this calculation in Table \ref{condi}, where $\Lambda_c$ is the
critical value for the self-trapping \cite{smerzi02} and $\Lambda$
is our solutions.

\begin{table}[h]
\caption{Different Numbers for the First Symmetric Breaking
Solutions} \label{condi} \centering
\begin{tabular}{|l|l|l|}
\hline\hline $ \eta$ & $\Lambda_c$ & $\Lambda$
\\ \hline
 5.0627 & 0.4945 & 0.9955
\\ \hline
 10.0627 & 0.4940 & 0.9960
 \\ \hline
 20.0627 & 0.4932 & 0.9985
\\
\hline
\end{tabular}
\end{table}

\begin{figure}[tbp]
\includegraphics[width=3.2 in]{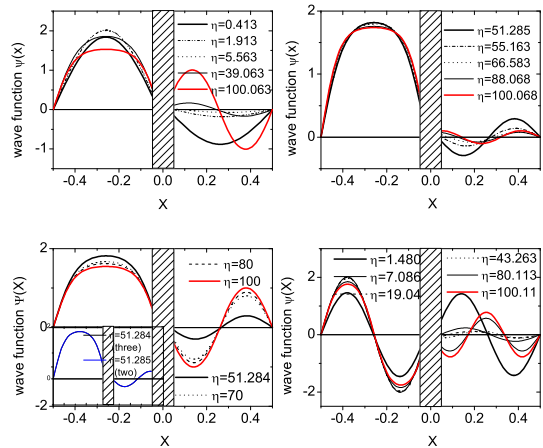}
\centering \vspace{-0.0 in} \caption{(Color online) Same as Fig.2
for negative nonlinear interaction} \label{fig3}
\end{figure}

In Fig.3, the symmetry breaking solution for negative nonlinear
interaction has been presented up to $-100$. When $\eta =-100$,
they are the same as the Fig. 4 in \cite{kutz01} but with a
different order. Same as the symmetry breaking solutions in
positive nonlinear interaction, the first and fourth solutions
emerge from the symmetry preserving solution at $\eta _c\sim
-0.32$ and at $\eta _c\sim -1.46$ respectively. Please note that
the first kind of symmetry breaking solution emerges from the
ground state and its chemical potential is less than that of the
ground state. Therefore this state is the ground state of the
system (From Fig.4). This is what is called the quantum phase
transition in this system \cite{kutz01,presilla01}. This solution
does not have any node and its profile show that the particle
would stay in one of the wells for negative nonlinear interaction.
The fourth solution includes one fixed node and a moving one. Now
the node moves from the well into the barrier region with the
decrease of the nonlinear interaction. Around $\eta \sim -33.239$
the node is in the barrier region.

Again, the second and the third solution occur from the isolated
point at $ \eta _c\sim -43.08$ in the diagram of the chemical
potential with the nonlinear interaction. Now their nodes become
fixed. Again the density profile has completely different behavior
with the decrease of the nonlinear interaction. The second
solution will collect the particles to one well but the third one
keeps them in two wells. There is still another crossover between
the third and the fourth solution near $-100$. That is why my
order of the solution is different from \cite{kutz01} at $\eta
=-100$.

\begin{figure}[tbp]
\includegraphics[width=3.2 in]{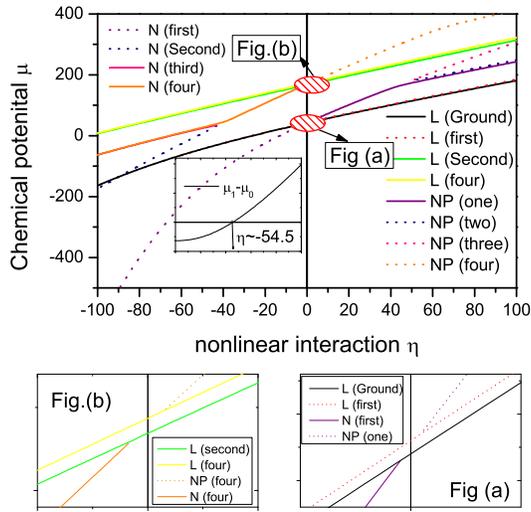}
\centering \vspace{-0.1 in} \caption{(Color online) The chemical
potential as the function of nonlinear interaction} \label{fig4}
\end{figure}

In Fig. 4. we plot the chemical potential as the function of
nonlinear interaction both for symmetry preserving and breaking
solutions. Due to the high barrier, the quantum tunneling effect
between the two wells is small and the  chemical potential
difference between the ground state and the first excited state is
small too. From Fig.4, one can directly see that the symmetry
preserving solution can be reduced to the eigenstate of the linear
Schr\"{o}dinger equations ($\eta =0$), while the symmetry breaking
one can not. To see this more clearly, we amplify the figure
around $\eta =0$ (Fig(a) is for the ground state and the first
excited state and Fig(b) is for the second excited state and the
third excited state). The embedded figure in Fig.4 (left) shows
the critical point for the quantum phase transition
\cite{kutz01,presilla01}. We denote the symmetry preserving
solutions as ``L'', and the symmetry breaking solutions and
positive nonlinear interaction as ``NP'', and the symmetry
breaking solutions and negative nonlinear interaction as ``N''. It
is clearly shown that the symmetry breaking solution can emerge
from two kind of points. One is the bifurcation and the other one
is the isolated point in this diagram. We arrange our solution in
Fig.1, Fig.2 and Fig.3 in the chemical potential order. It is easy
to see that the nature of the mean-field ground-state and the
structure of the energy spectrum of the nonlinear system depend on
the value of the nonlinear interaction.

\section{Conclusion}

We have presented one new analytical solution which could be
reduced to eigenstates of the corresponding linear Schr\"{o}dinger
equations. And we have shown the clear and direct evidence to the
relation of the symmetry preserving stationary solution and the
eigenstates of the linear Schr\"{o}dinger equations. Based on the
origin of the symmetry breaking state, one can find two kinds of
different symmetry breaking solutions, neither of which can be
reduced to the eigenstates of the corresponding linear
Schr\"{o}dinger equations and have one critical nonlinear
interaction, over which the symmetry breaking solution emerge. One
of the interesting things is that the first kind of symmetry
breaking solution can exist under the following conditions: 1. for
positive nonlinear interaction, it emerges from the first excited
state and the third excited state; 2. for negative nonlinear
interaction, it emerges from the ground state and the second
excited state. We understand this based on the superposition of
the eigenfunctions of the linear Schr\"{o}dinger equations. And
high accuracy of the critical value of the nonlinear interaction
has been found for the first kind of symmetry breaking solution
which emerges from the symmetry preserving one. This explains why
we can not find the symmetry breaking solution from the ground
state for the positive nonlinear interaction \cite{li}. As this
method is just valid for the small nonlinear interaction, we can
not yet understand the second kind of symmetry breaking solution.
But how the profile of the stationary solution change with the
nonlinear interaction has also been presented. Our calculation
shows that the symmetry breaking solution satisfies the
self-trapping condition. This will help to understand this effect
from stationary solutions.

Considering the situations realized in the experiment, our model
can be regarded as too simple. But this model can be solved
analytically and is good enough to provide qualitative
description. Usually, one can understand the realistic double well
by the quartic function $a x^4 + bx^2 +c$. The chemical potential
or the eigenvalue of GPE may show a little difference from the
results of our simple model, but they are consistent
quantitatively \cite{presilla01}.

WeiDong Li thanks the stimulative discussion with A. Smerzi and L.
Pitaevskii. This work is supported by the NSF of China (No.
10444002), SRF for ROCS, SEM of China and SRF for ROCS of the
Shanxi Province.

\emph{Note added.} Recently, we noted two papers \cite{zin,zin1}
which approach the same problem with similar techniques. In the
present work we provide a more detailed analysis of the density
profiles of stationary solutions as a function of the nonlinear
interaction and we find a symmetry-breaking solution.








\begin{thebibliography}{99}
\bibitem{sandro01} F. Dalfovo, S. Giorgini, L. P. Pitaevskii and S.
Stringari, Rev. of Mod. Phys. \textbf{71}, 463 (1999); A. J. Leggett, Rev.
of Mod. Phys. \textbf{73}, 307 (2001).

\bibitem{kutz01} K. W. Mahmud, J. N. Kutz and W. P. Reinhardt, Phys. Rev. A.
\textbf{66}, 063607 (2002).

\bibitem{presilla01} R. D'Agosta and C. Presilla, Phys. Rev. A. \textbf{65},
043609, (2002).

\bibitem{philip01} J. Denschlag, J. E. Simsarian et. al., Science \textbf{287%
} , 97 (2000).

\bibitem{burger01} S. Burger, K. Bongs et al., Phys. Rev. Lett. \textbf{83},
5198 (1999).

\bibitem{niu} Biao Wu, Jie Liu and Qian Niu. Phys. Rev. Lett. \textbf{88},
34101 (2002).

\bibitem{salomon01} L. Khaykovich, F. Schreck et. al., Science \textbf{292},
1290 (2002).

\bibitem{hulet01} K. E. Strecker, G. B. Partidge, A. G. Truscott and R. G.
Hulet, Nature, \textbf{417}, 150 (2002).

\bibitem{smerzi01} A. Smerzi, S. Fantoni, S. Giovanazzi and S. R. Shenoy,
Phys. Rev. Lett. \textbf{79}, 4950 (1997).

\bibitem{smerzi02} S. Raghavan, A. Smerzi, S. Fantoni and S. R. Shenoy,
Phys. Rev. A, \textbf{59} 620 (1999).

\bibitem{gati01} M. Albeiz, R. Gati, et al., cond-matt/0411757.

\bibitem{gati02} Th. Anker, M. Albiez, R. Gati, S. Hunsmann, B. Eiermann, A.
Trombettoni and M. K. Oberthaler, Phys. Rev. Lett. \textbf{94}, 20403 (2005).

\bibitem{pethick01} M. Machholm, A. Nicolin,  C. J. Pethick and H.
Smith, Phys. Rev. A, \textbf{69} 043604 (2004).

\bibitem{smerzi03}  W. D. Li and A. Smerzi, Phys. Rev. E. \textbf{70}
16605 (2004).

\bibitem{gati03} B. Eiermann, T. Anker, M. Albiez, M. Taglieber, P.
Treutlein, K. -P. Marzlin and M. K. Oberthaler, Phys. Rev. Lett. \textbf{92}%
, 230401(2004).

\bibitem{byrd01} P. F. Byrd, M. D. Driedman, Handbook of Elliptic Integrals
for Engineers and Scientists, Second Edition, Springer-Verlage New York
Heidelberg Berlin 1971.

\bibitem{li} XinYan Jia, WeiDong Li and Hiroshi Ezawa, in preparing.

\bibitem{zin} P. Zin, E. Infeld, M. Matuszewski, G. Rowlands and M.
Trippenbach, Phys. Rev. A. {\bf{73}} 022105 (2006).

\bibitem{zin1} E. Infeld, P. Zin, J. Gocalik, and M. Trippenbach,
Phys. Rev. E \textbf{74}, 026610 (2006).
\end{thebibliography}
\end{document}